\documentstyle[12pt,axodraw]{article}

\catcode`@=12
\begin{document}
\thispagestyle{empty}
\begin{flushright} 
UCRHEP-T300\\ 
March 2001\
\end{flushright}
\vspace{0.5in}
\begin{center}
{\LARGE	\bf Making Neutrinos Massive with\\ an Axion in Supersymmetry\\}
\vspace{1.5in}
{\bf Ernest Ma\\}
\vspace{0.2in}
{\sl Physics Department, University of California, Riverside, 
California 92521\\}
\vspace{1.5in}
\end{center}
\begin{abstract}\
The minimal supersymmetric standard model (MSSM) of particle interactions is 
extended to include three singlet (right-handed) neutrino superfields 
together with three other singlet superfields.  The resulting theory is 
assumed to be invariant under an anomalous global U(1) (Peccei-Quinn) 
symmetry with one fundamental mass $m_2$.  The soft breaking of supersymmetry 
at the TeV scale is shown to generate an axion scale $f_a$ of order $m_2$. 
Neutrino masses are generated by $f_a$ according to the usual seesaw mechanism.
\end{abstract}
\newpage
\baselineskip 24pt

The minimal supersymmetric standard model (MSSM) is a well-motivated extension 
of the standard model (SM) of particle interactions.  Nevertheless, it is 
missing at least two important ingredients.  There is no neutrino mass and 
the strong CP problem \cite{cpv} is unresolved.  Whereas separate remedies 
exist for both shortcomings, they are in general unrelated \cite{prev}. 
In the following, I start with a supersymmetric theory of just one large 
fundamental mass $(m_2)$.  I assume it to be invariant under an anomalous 
global U(1) symmetry which is an extension of the well-known Peccei-Quinn 
symmetry \cite{pq} to include three singlet (right-handed) neutrino 
superfields $(\hat N^c)$ and three other singlet superfields $(\hat S)$.  
The supersymmetry is then softly broken at $M_{SUSY}$ of order 1 TeV.  As 
a result of the assumed particle content of the theory, an axion scale $f_a$ 
of order $m_2$ is generated, from which neutrinos obtain masses 
via the usual seesaw mechanism with $m_N \sim f_a$. 

Consider first the MSSM superpotential:
\begin{equation}
\hat W = \mu \hat H_u \hat H_d + h_u \hat H_u \hat Q \hat u^c + h_d \hat H_d 
\hat Q \hat d^c + h_e \hat H_d \hat L \hat e^c.
\end{equation}
Under $U(1)_{PQ}$, the quark $(\hat Q, \hat u^c, \hat d^c)$ and lepton 
$(\hat L, \hat e^c)$ superfields have charges $+1/2$, whereas the Higgs 
$(\hat H_u, \hat H_d)$ superfields have charges $-1$.  Hence the $\mu$ 
term is forbidden.  It is replaced here by $h_2 \hat S_2 \hat H_u 
\hat H_d$, where $\hat S_2$ is a singlet superfield with PQ charge $+2$.

Add three singlet superfields $\hat N^c$ with PQ charges $+1/2$.  The 
term $h_N \hat H_u \hat L \hat N^c$ is then allowed, but the usual 
Majorana mass term $m_N \hat N^c \hat N^c$ is forbidden.  Instead, it is 
replaced by $h_1 \hat S_1 \hat N^c \hat N^c$, where $\hat S_1$ has PQ 
charge $-1$.

So far there is no mass scale in the superpotential of this theory.  It is 
thus natural to introduce a third singlet superfield $\hat S_0$ with PQ 
charge $-2$ so that the complete superpotential of this theory is given by
\begin{eqnarray}
\hat W &=& m_2 \hat S_2 \hat S_0 + f \hat S_1 \hat S_1 \hat S_2 + h_1 \hat S_1 
\hat N^c \hat N^c + h_2 \hat S_2 \hat H_u \hat H_d \nonumber \\ &+& 
h_N \hat H_u \hat L \hat N^c + h_u \hat H_u \hat Q \hat u^c + h_d \hat H_d 
\hat Q \hat d^c + h_e \hat H_d \hat L \hat e^c.
\end{eqnarray}
The mass $m_2$ is a large fundamental scale which will be shown to coincide 
with the axion scale, even though supersymmetry is only broken at the TeV 
scale.  The term $\hat S_1 \hat S_1 \hat S_2$ is automatically present 
and will be the key to understanding how $f_a$ is generated from $M_{SUSY}$. 
Consider next the spontaneous breaking of $U(1)_{PQ}$ by the vacuum 
expectation values $v_{2,1,0}$ of $S_{2,1,0}$ respectively.  The $\mu$ term 
of the MSSM is then given by $h_2 v_2$ and the Majorana mass of the neutrino 
singlet is $2 h_1 v_1$.  Hence $v_1$ should be many orders of magnitude 
greater than $v_2$.  With $m_N = 2 h_1 v_1$, the usual seesaw relationship
\begin{equation}
m_\nu = {h_N^2 v_u^2 \over m_N}
\end{equation}
is also obtained.  Now the axion scale $f_a$ is of order $v_1$ as well, thus 
$m_N \sim f_a$.  Whereas this relationship was also proposed previously 
\cite{prev}, the hierarchy problem of $v_2 << v_1$ was not addressed.  If a 
sterile neutrino superfield $\hat \nu_s$ is desirable, it may be assigned PQ 
charge $-5/2$.  The term $h_s \hat S_2 \hat \nu_s \hat N^c$ is then allowed 
in Eq.~(2), resulting in a seesaw mass for $\nu_s$ given by $h_s^2 v_2^2/m_N$.

The strong CP problem is the problem of having the instanton-induced term 
\cite{cpv}
\begin{equation}
{\cal L}_\theta = \theta_{QCD} {g_s^2 \over 64 \pi^2} \epsilon_{\mu \nu 
\alpha \beta} G_a^{\mu \nu} G_a^{\alpha \beta}
\end{equation}
in the effective Lagrangian of quantum chromodynamics (QCD), where $g_s$ 
is the strong coupling constant, and
\begin{equation}
G_a^{\mu \nu} = \partial^\mu G_a^\nu - \partial^\nu G_a^\mu + g_s f_{abc} 
G_b^\mu G_c^\nu
\end{equation}
is the gluonic field strength.  If $\theta_{QCD}$ is of order unity, the 
neutron electric dipole moment is expected \cite{edm} to be $10^{10}$ 
times its present experimental upper limit ($0.63 \times 10^{-25}~e$ cm) 
\cite{nedm}.  This conundrum is most elegantly resolved by invoking 
a dynamical mechanism \cite{pq} to relax the above $\theta_{QCD}$ parameter 
(including all contributions from colored fermions) to zero.  However, this 
necessarily results \cite{ww} in a very light pseudoscalar particle called 
the axion, which has not yet been observed \cite{search}.

To reconcile the nonobservation of an axion in present experiments and the 
constraint $10^9 ~{\rm GeV} < f_a < 10^{12}$ GeV from astrophysics and 
cosmology \cite{astro}, three types of 
``invisible'' axions have been discussed.  The DFSZ solution \cite{dfsz} 
introduces a heavy singlet scalar field as the source of the axion but 
its mixing with the doublet scalar fields (which couple to the usual quarks) 
is very much suppressed.  The KSVZ solution \cite{ksvz} also has a heavy 
singlet scalar field but it couples only to new heavy colored fermions.
The gluino solution \cite{dema} identifies the $U(1)_R$ of superfield 
transformations with $U(1)_{PQ}$ and thus the axion is a dynamical phase 
attached to the gluino as well as all other superparticles.  The present 
model is of the DFSZ type, but the use of 3 fundamental superfields is 
motivated by Refs.\cite{dema2,demasa}.

Before discussing the spontaneous breaking of $U(1)_{PQ}$ in the context of 
soft supersymmetry breaking, consider 
$\hat W$ of Eq.~(2) in terms of baryon number and lepton number.  It is 
clear that the former is conserved as a global symmetry ($\hat Q$ has $B = 
1/3$, $\hat u^c$ and $\hat d^c$ have $B = -1/3$, all others have $B = 0$), 
whereas the latter is conserved only as a discrete symmetry ($\hat L$, 
$\hat e^c$, and $\hat N^c$ are odd, all others are even).  Thus 
the usual $R$ parity of the MSSM is also conserved.  The three $\hat N^c$ 
superfields are well-motivated because they allow small seesaw neutrino 
masses for neutrino oscillations \cite{atm,sol,lsnd}. 
The Peccei-Quinn symmetry is well-motivated as the most attractive solution 
of the strong CP problem.  Hence $S_1$ and $S_2$ are both well-motivated. 
Finally, $S_0$ is well-motivated because $\hat W$ should have a large 
fundamental mass scale.  Given all these well-motivated inputs, Eq.~(2) is 
uniquely determined and the two crucial extra terms $m_2 \hat S_2 \hat S_0$ 
and $f \hat S_1 \hat S_1 \hat S_2$ are automatically present \cite{adhoc}.

Consider the scalar potential of $S_{2,1,0}$, i.e.
\begin{equation}
V = |m_2 S_0 + f S_1^2|^2 + m_2^2 |S_2|^2 + 4 f^2 |S_1|^2 |S_2|^2.
\end{equation}
There are two supersymmetric minima: the trivial one with $v_0 = v_1 = 
v_2 = 0$, and the much more interesting one with
\begin{equation}
v_2 = 0, ~~~ m_2 v_0 + f v_1^2 = 0,
\end{equation}
where $v_{2,1,0} = \langle S_{2,1,0} \rangle$.  The latter breaks 
$U(1)_{PQ}$ spontaneously and the superpotential of $\hat S_{2,1,0}$ becomes
\begin{eqnarray}
\hat W' = {m_2 \over v_1} (v_1 \hat S_0 - 2 v_0 \hat S_1) \hat S_2 + 
f \hat S_1 \hat S_1 \hat S_2,
\end{eqnarray}
after shifting by $v_{2,1,0}$. This shows clearly that the linear combination
\begin{equation}
{v_1 \hat S_1 + 2 v_0 \hat S_0 \over \sqrt {|v_1|^2 + 4 |v_0|^2}}
\end{equation}
is a massless superfield.  Hence the axion is not even contained in 
$\hat S_2$, and its effective coupling through $\hat S_2$ will be 
suppressed by $M_{SUSY}/v_{1,0}$ as desired.

At this point, the individual values of $v_1$ and $v_0$ are not determined. 
This is because the vacuum is invaraint not only under a phase rotation but 
also under a scale transformation as a result of the unbroken supersymmetry 
\cite{spont}.  As such, it is unstable and the soft breaking of supersymmetry 
at the TeV scale will determine $v_1$ and $v_0$, and $v_2$ will become 
nonzero.  Specifically, the supersymmetry of this theory is assumed broken 
by all possible holomorphic soft terms which are invariant under $U(1)_{PQ}$. 
In particular, all the usual MSSM soft terms are present except for the 
$\mu B H_u H_d$ term.  However, there is the $h_2 A_2 S_2 H_u 
H_d$ term as well as the $|m_2 S_0 + f S_1 S_1 + h_2 H_u H_d|^2$ term, 
hence ``$\mu B$'' = $h_2 [A_2 v_2 + (m_2 v_0 + f v_1^2)]$.  Recall that the 
$\mu$ parameter of the MSSM is replaced here by $h_2 v_2$.  Hence $v_2$ 
should be of order $M_{SUSY}$ and $m_2 v_0 + f v_1^2$ of order $M_{SUSY}^2$, 
and that is exactly what will be shown in the following.

Add now the other holomorphic soft terms of the scalar potential:
\begin{eqnarray}
V_{soft} &=& \mu_0^2 |S_0|^2 + \mu_1^2 |S_1|^2 + \mu_2^2 |S_2|^2 \nonumber \\ 
&+& [\mu_{20} m_2 S_2 S_0 + \mu_{12} S_1^2 S_2 + h.c.],
\end{eqnarray}
where all new parameters are assumed of order $M_{SUSY} \sim 1$ TeV.  
Consider the minimum of the scalar potential of $S_{2,1,0}$ with the 
addition of $V_{soft}$, i.e.
\begin{eqnarray}
V_{min} &=& (m_2^2 + \mu_0^2) v_0^2 + \mu_1^2 v_1^2 + (m_2^2 + \mu_2^2) v_2^2 
+ 2 \mu_{20} m_2 v_2 v_0 \nonumber \\ &+& 2 m_2 f v_1^2 v_0 + 
2 \mu_{12} v_1^2 v_2 + 4 f^2 v_2^2 v_1^2 + f^2 v_1^4,
\end{eqnarray}
where every quantity has been assumed real for simplicity.  The conditions 
on $v_{2,1,0}$ are
\begin{eqnarray}
&& (m_2^2 + \mu_2^2 + 4 f^2 v_1^2) v_2 + \mu_{20} m_2 v_0 + \mu_{12} v_1^2 = 
0, \\ && f(m_2 v_0 + f v_1^2) + {1 \over 2} \mu_1^2 + \mu_{12} v_2 + 
2 f^2 v_2^2 = 0, \\ && m_2 (m_2 v_0 + f v_1^2) + \mu_0^2 v_0 + \mu_{20} 
m_2 v_2 = 0.
\end{eqnarray}
Together they show that Eq.~(7) is modified to read
\begin{equation}
v_2 \sim M_{SUSY}, ~~~ m_2 v_0 + f v_1^2 \sim M_{SUSY}^2.
\end{equation}
However, $v_0$ and $v_1$ are individually of order $m_2$.  This can be 
shown by integrating out the heavy fields $S_{2,0}$ and $\tilde S_{2,0}$ 
in the limit where $S_1$ is massless.  The effective scalar potential for 
$S_1$ is necessarily of the form
\begin{equation}
V_{eff} = a |S_1|^2 + b |S_1|^4.
\end{equation}
At tree level, $a = \mu_1^2$, and 
\begin{equation}
b = f^2 - {(m_2 f)^2 \over m_2^2 + \mu_0^2} - {\mu_{12}^2 \over m_2^2 + 
\mu_2^2} \simeq {f^2 \mu_0^2 - \mu_{12}^2 \over m_2^2},
\end{equation}
where the mixing parameter $\mu_{20}$ has been neglected.  Hence
\begin{equation}
v_1^2 = {-a \over 2b} \simeq {-\mu_1^2 m_2^2 \over 2 (f^2 \mu_0^2 - 
\mu_{12}^2)}.
\end{equation}
Since all soft supersymmetry breaking parameters are of order $M_{SUSY}$, 
this shows that $v_1$ is of order $m_2$ and so is $v_0$.  The parameters 
$a$ and $b$ have logarithmically divergent corrections in one loop, but they 
are proportional to $M_{SUSY}^2$ and $M_{SUSY}^2/m_2^2$ respectively, hence 
they do not spoil the tree-level result of Eq.~(18).  [This would not be the 
case if the nonholomorphic soft term $S_1^2 S_0^*$ were added.]

To understand why $v_1 >> M_{SUSY}$ is possible, consider the superfield 
of Eq.~(9).  The phase of the 
corresponding scalar field is the axion, but the magnitude is a physical 
scalar particle of mass $\sim M_{SUSY}$, and the associated Majorana 
fermion (axino) also has a mass $\sim M_{SUSY}$.  Hence $v_1 \neq 0$ does 
not necessarily imply that supersymmetry is broken at that scale.  For 
example, the superfield $N^c$ has the large mass $2 h_1 v_1$, and $M_{SUSY}$ 
accounts only for the relative small splitting between its scalar and fermion 
components.

As the electroweak $SU(2)_L \times U(1)_Y$ gauge symmetry is broken by the 
vacuum expectation values $v_{u,d}$ of $H_{u,d}$, the observed doublet 
neutrinos acquire naturally small Majorana masses given by $m_\nu = h_N^2 
v_u^2 /(2 f v_1)$ via the usual seesaw mechanism.  Since $H_{u,d}$ have 
PQ charges as well, the axion field is now given by
\begin{equation}
a = {v_1 \theta_1 + 2 v_0 \theta_0 - 2 v_2 \theta_2 + v_u \theta_u 
+ v_d \theta_d \over V}
\end{equation}
where $V = \sqrt {v_1^2 + 4v_0^2 + 4v_2^2 + v_u^2 + v_d^2}$, and 
$\theta_i$ are the various properly normalized angular fields, 
from the decomposition of a complex scalar field $\phi = (1/\sqrt 2) (v + 
\rho) \exp (i\theta/v)$ with the kinetic energy term
\begin{equation}
\partial_\mu \phi^* \partial^\mu \phi = {1 \over 2} (\partial_\mu \rho)^2 
+ {1 \over 2} (\partial_\mu \theta)^2 \left( 1 + {\rho \over v} \right)^2.
\end{equation}
The axionic coupling to quarks is thus
\begin{equation}
(\partial_\mu a) \left[ {1 \over 2} \left( {v_u \over V} \right) {1 \over 
v_u} \bar u \gamma^\mu \gamma_5 u + {1 \over 2} \left( {v_d \over V} 
\right) {1 \over v_d} \bar d \gamma^\mu \gamma_5 d \right] = {1 \over 2 V} 
(\partial_\mu a) \sum_{q=u,d} \bar q \gamma^\mu \gamma_5 q,
\end{equation}
as in the DFSZ model.

Consider now all the physical particles of this theory.  (1) There is a 
heavy Dirac fermion of mass $m_2 \sqrt {1+4v_0^2/v_1^2}$, formed out of 
$\tilde S_2$ and a linear combination of $\tilde S_0$ and $\tilde S_1$.  
The two associated scalars also have the same mass but with vacuum expectation 
values of order $M_{SUSY}$.  (2) There are three heavy $N^c$ 
superfields with mass of order $\langle S_1 \rangle = v_1 \sim m_2$.  
They provide seesaw masses for the neutrinos and generate a 
primordial lepton asymmetry through their decays \cite{fuya}.  This gets 
converted into the present observed baryon asymmetry of the Universe 
through the $B+L$ violating electroweak sphalerons \cite{kurush}.  (3) 
The particles of the MSSM and their interactions are all present, but with 
the $\mu$ parameter given by $h_2 v_2 \sim M_{SUSY}$ and the $\mu B$ parameter 
by $h_2 [A_2 v_2 + (m_2 v_0 + f v_1^2)] \sim M_{SUSY}^2$, thus solving the 
$\mu$ problem (i.e. why $\mu \sim M_{SUSY}$ and not $m_2$) without causing a 
$\mu B$ problem.  (4) Whereas the spontaneous breaking of $U(1)_{PQ}$ 
generates an axion at the scale $V \sim m_2$, thus solving the strong CP 
problem, the physical scalar field (saxion) with this dynamical phase has 
a mass $\sim M_{SUSY}$.  It is effectively unobservable \cite{dema2} because 
its couplings to all MSSM particles are suppressed by at least $v_{2,u,d}/V$.  
(5) An axino of mass $\sim M_{SUSY}$ also exists.  Since $R$ parity is 
conserved and the axino has $R = -1$, it may be the stable LSP (lightest 
supersymmetric particle) of this theory and be experimentally observed. 
\cite{marasa} 

The $7 \times 7$ mass matrix spanning the $R = -1$ neutral fermions of this 
theory in the basis $(\tilde B, \tilde W_3, \tilde H_u^0, \tilde H_d^0, \tilde 
S_2, \tilde S_0, \tilde S_1)$ is given by
\begin{equation}
{\cal M} = \left[ \begin{array}
{c@{\quad}c@{\quad}c@{\quad}c@{\quad}c@{\quad}c@{\quad}c}
\tilde m_1 & 0 & -sm_3 & sm_4 & 0 & 0 & 0 \\ 
0 & \tilde m_2 & cm_3 & -cm_4 & 0 & 0 & 0 \\ 
-sm_3 & cm_3 & 0 & h_2 v_2 & h_2 v_d & 0 & 0 \\ 
sm_4 & -cm_4 & h_2 v_2 & 0 & h_2 v_u & 0 & 0 \\ 
0 & 0 & h_2 v_d & h_2 v_u & 0 & m_2 & 2fv_1 \\ 
0 & 0 & 0 & 0 & m_2 & 0 & 0 \\ 
0 & 0 & 0 & 0 & 2fv_1 & 0 & 2 f v_2 \end{array} \right],
\end{equation}
where $s = \sin \theta_W$, $c = \cos \theta_W$, $m_3 = M_Z \cos \beta$, 
$m_4 = M_Z \sin \beta$, with $\tan \beta = v_u/v_d$.  Without $\tilde 
S_{2,1,0}$, the above is just the neutralino mass matrix of the MSSM and 
the LSP is a linear combination of the two gauginos and the two Higgsinos.  
In this theory, that combination has a small overlap with the axino of order 
$v_{u,d}/V$.  If kinematically allowed, it will decay into the axino and 
the $Z$ boson or a neutral Higgs boson.

In conclusion, a desirable extension of the MSSM has been presented which 
has only two input scales, i.e. the large fundamental scale $m_2$ and the 
soft supersymmetry breaking scale $M_{SUSY}$.  Assuming the validity of 
$U(1)_{PQ}$ and its implementation in terms of Eqs.~(2) and (10), an axion 
scale $\sim m_2$ is generated, which solves the strong CP 
problem \underline {and} makes neutrinos massive via the usual seesaw 
mechanism, \underline {without} breaking the supersymmetry at $m_2$.  The 
baryon asymmetry of the Universe is accommodated as well as 
the existence of dark matter.  The $\mu$ problem of the MSSM is solved 
without causing a $\mu B$ problem.  The other particles associated with 
the axion (saxion and axino) have masses of order $M_{SUSY}$, with the 
axino as a candidate for the true lightest supersymmetric particle.\\[5pt]

I thank G. Senjanovic and F. Zwirner for very helpful discussions. 
This work was supported in part by the U.~S.~Department of Energy
under Grant No.~DE-FG03-94ER40837.

\newpage
\bibliographystyle{unsrt}

\begin{thebibliography}{99}

\bibitem{cpv}
C.~G.~Callan, R.~F.~Dashen, and D.~J.~Gross,
%``The structure of the gauge theory vacuum,''
Phys.\ Lett.\ {\bf B63}, 334 (1976);
%%CITATION = PHLTA,B63,334;%%
R.~Jackiw and C.~Rebbi,
%``Vacuum periodicity in a Yang-Mills quantum theory,''
Phys.\ Rev.\ Lett.\ {\bf 37}, 172 (1976).
%%CITATION = PRLTA,37,172;%%

\bibitem{prev} There have been exceptions in nonsupersymmetric models. 
See for example R. N. Mohapatra and G. Senjanovic, Z. Phys. {\bf C17}, 53 
(1983); P. Langacker, R. D. Peccei, and T. Yanagida, Mod. Phys. Lett. 
{\bf A1}, 541 (1986); M. Shin, Phys. Rev. Lett. {\bf 59}, 2515 (1987); 
Erratum: {\bf 60}, 383 (1988).

\bibitem{pq} 
R.~D.~Peccei and H.~R.~Quinn,
%``CP Conservation In The Presence Of Instantons,''
Phys.\ Rev.\ Lett.\ {\bf 38}, 1440 (1977).
%%CITATION = PRLTA,38,1440;%%

\bibitem{edm} 
V.~Baluni,
%``CP Violating Effects In QCD,''
Phys.\ Rev.\ D {\bf 19}, 2227 (1979);
%%CITATION = PHRVA,D19,2227;%%
R.~J.~Crewther, P.~Di Vecchia, G.~Veneziano, and E.~Witten,
%``Chiral Estimate Of The Electric Dipole Moment Of The Neutron In Quantum Chromodynamics,''
Phys.\ Lett.\ {\bf B88}, 123 (1979).
%%CITATION = PHLTA,B88,123;%%

\bibitem{nedm} 
P.~G.~Harris {\it et al.},
%``New experimental limit on the electric dipole moment of the neutron,''
Phys.\ Rev.\ Lett.\ {\bf 82}, 904 (1999).
%%CITATION = PRLTA,82,904;%%

\bibitem{ww} 
S.~Weinberg,
%``A New Light Boson?,''
Phys.\ Rev.\ Lett.\ {\bf 40}, 223 (1978);
%%CITATION = PRLTA,40,223;%%
F.~Wilczek,
%``Problem Of Strong P And T Invariance In The Presence Of Instantons,''
Phys.\ Rev.\ Lett.\ {\bf 40}, 279 (1978).
%%CITATION = PRLTA,40,279;%%

\bibitem{search} 
L.~J.~Rosenberg and K.~A.~van Bibber,
%``Searches for invisible axions,''
Phys.\ Rept.\ {\bf 325}, 1 (2000).
%%CITATION = PRPLC,325,1;%%

\bibitem{astro} 
G.~G.~Raffelt,
%``Particle physics from stars,''
Ann.\ Rev.\ Nucl.\ Part.\ Sci.\ {\bf 49}, 163 (1999)
[hep-ph/9903472].
%%CITATION = HEP-PH 9903472;%%

\bibitem{dfsz}
M.~Dine, W.~Fischler, and M.~Srednicki,
%``A Simple Solution To The Strong CP Problem With A Harmless Axion,''
Phys.\ Lett.\ {\bf B104}, 199 (1981);
%%CITATION = PHLTA,B104,199;%%
A.~R.~Zhitnitsky,
%``On Possible Suppression Of The Axion Hadron Interactions. (In Russian),''
Sov.\ J.\ Nucl.\ Phys.\ {\bf 31}, 260 (1980).
%%CITATION = SJNCA,31,260;%%

\bibitem{ksvz} 
J.~E.~Kim,
%``Weak Interaction Singlet And Strong CP Invariance,''
Phys.\ Rev.\ Lett.\ {\bf 43}, 103 (1979).
%%CITATION = PRLTA,43,103;%%
M.~A.~Shifman, A.~I.~Vainshtein, and V.~I.~Zakharov,
%``Can Confinement Ensure Natural CP Invariance Of Strong Interactions?,''
Nucl.\ Phys.\ {\bf B166}, 493 (1980).
%%CITATION = NUPHA,B166,493;%%

\bibitem{dema} 
D.~A.~Demir and E.~Ma,
%``Relaxation of the dynamical gluino phase and unambiguous electric  dipole moments,''
Phys.\ Rev.\ D {\bf 62}, 111901(R) (2000)
[hep-ph/0004148].
%%CITATION = HEP-PH 0004148;%%

\bibitem{dema2} D. A. Demir and E. Ma, hep-ph/0101185.

\bibitem{demasa} 
D.~A.~Demir, E.~Ma, and U.~Sarkar,
%``Neutrino masses and the gluino axion model,''
J.\ Phys.\ {\bf G26}, L117 (2000)
[hep-ph/0005288].
%%CITATION = HEP-PH 0005288;%%

\bibitem{atm} Y. Fukuda {\it et al.}, Super-Kamiokande Collaboration, 
Phys. Lett. {\bf B433}, 9 (1998); {\bf B436}, 33 (1998); {\bf B467}, 185 
(1999); Phys. Rev. Lett. {\bf 81}, 1562 (1998), {\bf 82}, 2644 (1999).

\bibitem{sol} Y. Fukuda {\it et al.}, Super-Kamiokande Collaboaration, 
Phys. Rev. Lett. {\bf 81}, 1158 (1998); {\bf 82}, 1810, 2430 (1999).

\bibitem{lsnd} C. Athanassopoulos {\it et al.}, Phys. Rev. Lett. {\bf 75}, 
2650 (1995); {\bf 77}, 3082 (1996); {\bf 81}, 1774 (1998).

\bibitem{adhoc} In Refs.\cite{dema2,demasa}, somewhat different structures 
of the three $\hat S$ superfields were used.  However, $\hat S_1$ served no 
other purpose and an extra {\it ad hoc} $Z_3$ discrete symmetry was needed. 
In addition, fine tuning of the effective scalar potential for $S_1$ is 
required at the one-loop level.

\bibitem{spont} 
E.~Ma,
%``Spontaneous supersymmetric generation of an indeterminate mass scale  and a possible light sterile neutrino,''
Mod.\ Phys.\ Lett.\ {\bf A14}, 1637 (1999)
[hep-ph/9904429].
%%CITATION = HEP-PH 9904429;%%

\bibitem{fuya} M. Fukugita and T. Yanagida, Phys. Lett. {\bf 174B}, 45 (1986).

\bibitem{kurush} V. A. Kuzmin, V. A. Rubakov, and M. E. Shaposhnikov, Phys. 
Lett. {\bf 155B}, 36 (1985).

\bibitem{marasa} In a recently proposed axion model with associated particles 
observable at the TeV scale, an anomalous gauge symmetry is required: 
E.~Ma, M.~Raidal, and U.~Sarkar, Phys. Lett. {\bf B}, in press 
%``Low-scale axion from large extra dimensions,''
[hep-ph/0007321].
%%CITATION = HEP-PH 0007321;%%


\end{thebibliography}

\end{document}